\documentclass[conference]{IEEEtran}
\IEEEoverridecommandlockouts
\usepackage{amsmath,amssymb,amsfonts}
\usepackage{algorithmic}
\usepackage{graphicx}
\usepackage{textcomp}
\usepackage{xcolor}
\usepackage{booktabs}
\usepackage{float}
\usepackage{multicol}
\usepackage{multirow}
\usepackage{afterpage}
\usepackage{subcaption}
\usepackage{threeparttable}

\usepackage[backend=biber, style=ieee]{biblatex}
\begin{document}

\title{A Tree-based RAG-Agent Recommendation System: A Case Study in Medical Test Data}

\author{\IEEEauthorblockN{Yahe Yang}
\IEEEauthorblockA{\textit{George Washington University}\\
yahe.yang@gwu.edu}

~\\
\and
\IEEEauthorblockN{Chengyue Huang}
\IEEEauthorblockA{\textit{University of Iowa}\\
chengyue-huang@uiowa.edu}
\textsuperscript{*}Corresponding author}

\maketitle

\begin{abstract}
We present HiRMed (Hierarchical RAG-enhanced Medical Test Recommendation), a novel tree-structured recommendation system that leverages Retrieval-Augmented Generation (RAG) for intelligent medical test recommendations. Unlike traditional vector similarity-based approaches, our system performs medical reasoning at each tree node through a specialized RAG process. Starting from the root node with initial symptoms, the system conducts step-wise medical analysis to identify potential underlying conditions and their corresponding diagnostic requirements. At each level, instead of simple matching, our RAG-enhanced nodes analyze retrieved medical knowledge to understand symptom-disease relationships and determine the most appropriate diagnostic path. The system dynamically adjusts its recommendation strategy based on medical reasoning results, considering factors such as urgency levels and diagnostic uncertainty. Experimental results demonstrate that our approach achieves superior performance in terms of coverage rate, accuracy, and miss rate compared to conventional retrieval-based methods. This work represents a significant advance in medical test recommendation by introducing medical reasoning capabilities into the traditional tree-based retrieval structure.
\end{abstract}


\section{Introduction}

Medical test recommendation plays a pivotal role in modern healthcare systems, directly influencing diagnostic accuracy, treatment outcomes, and resource utilization. While traditional approaches like rule-based systems and similarity-based retrieval methods have been widely used, they often fall short in capturing the nuanced, context-dependent nature of medical diagnosis. The challenge lies not only in recommending appropriate diagnostic tests based on patient symptoms but also in considering the complex interplay of medical priorities, resource constraints, and diagnostic uncertainty.

Recent advances in large language models (LLMs) and retrieval-augmented generation (RAG) have shown promising results in various healthcare applications. However, their direct application to medical test recommendations faces several critical limitations. These include the lack of hierarchical reasoning structure that mirrors medical diagnostic processes, insufficient integration of domain-specific medical knowledge at different diagnostic stages, limited ability to maintain and utilize reasoning history throughout the recommendation process, and the challenge in balancing between comprehensive coverage and targeted specificity in test recommendations.

To address these limitations, we propose HiRMed (Hierarchical RAG-enhanced Medical Test Recommendation), a novel system that combines hierarchical reasoning structures with RAG-enhanced medical decision support. Our approach is distinguished by three key innovations. First, our Hierarchical RAG Architecture implements a tree-structured architecture where each node incorporates specialized RAG processes, enabling progressive refinement of recommendations through multiple levels of medical reasoning, from general symptom assessment to specific test selection. Second, our system features Dynamic Knowledge Integration with a dual-layer knowledge base architecture - a root-level knowledge base for broader medical understanding and department-specific knowledge bases for specialized diagnostic considerations. This structure enables more precise and context-aware recommendations at different stages of the diagnostic process. Third, HiRMed incorporates Memory-Augmented Reasoning with a sophisticated memory mechanism that maintains reasoning history across nodes, enabling coherent and comprehensive diagnostic paths while avoiding redundant or conflicting recommendations.

The significance of our work extends beyond theoretical innovation. Through comprehensive empirical evaluation, we demonstrate that HiRMed achieves higher accuracy in test recommendations compared to traditional approaches, improved coverage of potential diagnostic paths, lower miss rates for critical diagnostic tests, and better interpretability of recommendation rationale through explicit reasoning paths. This paper presents both the theoretical framework underpinning HiRMed and its practical implementation, providing a blueprint for next-generation medical test recommendation systems that combine the power of LLMs with structured medical reasoning.

\section{Related Work}

The intersection of artificial intelligence and healthcare has seen significant advances, particularly in the realm of recommendation systems for diagnostic and treatment purposes. This section reviews existing work in tree-based recommendation systems, Retrieval-Augmented Generation (RAG), and their applications in medical diagnostics.

Tree-based recommendation systems have long been a cornerstone of decision support systems, leveraging hierarchical structures to mirror human decision-making processes. For instance, hierarchical models have been applied to stratify patient symptoms and align them with diagnostic categories \cite{Kotov2018A,Gara1998A,Gilam2021Classifying}. These models have shown promise in reducing the complexity of diagnostic workflows by breaking them into manageable, structured layers. These systems often rely on static rule-based approaches, limiting their adaptability to dynamic medical data and evolving clinical guidelines \cite{Fryan2022Processing,Zhang2023Decision}.

Retrieval-Augmented Generation (RAG) has emerged as a transformative approach in natural language processing, combining retrieval mechanisms with generative models. Lewis et al. \cite{lewis2020retrieval} introduced RAG to enhance generative models by incorporating retrieved documents, demonstrating improvements in tasks requiring factual grounding. This approach excels in providing evidence-based answers in open-domain question answering, where retrieved contextual information significantly improves response quality. While RAG has been explored in such contexts, its application to structured, domain-specific tasks like medical diagnostics remains underexplored. Notable exceptions include Miao et al. \cite{Miao2024Integrating}, who used RAG for symptom-to-disease mapping but lacked a hierarchical structure to emulate the structured reasoning essential in medical diagnostics.

In the context of medical test recommendations, prior studies have predominantly utilized similarity-based methods. Traditional approaches match patient symptoms to diagnostic tests using vector similarity \cite{Jia2019A,Sharafoddini2017Patient,Parimbelli2018Patient}. While effective in straightforward cases, these methods may have limits in incorporating nuanced medical reasoning, leading to oversights in complex diagnostic scenarios. 

Recent advancements in hierarchical reasoning systems offer promising alternatives. For example, Zhang et al. \cite{zhang2022drlk} proposed dynamic hierarchical interactions between the QA context and knowledge graph for reasoning, emphasizing the importance of maintaining a reasoning history. Their work demonstrated that hierarchical structures not only enhance interpretability but also provide a mechanism for systematically refining diagnostic hypotheses. This aligns closely with the objectives of HiRMed, which incorporates a memory-augmented reasoning mechanism to ensure coherence across diagnostic steps. By retaining context from earlier stages of the recommendation process, such systems mitigate the risk of inconsistent or fragmented diagnostic paths.

Furthermore, domain-specific knowledge integration remains a critical challenge in medical recommendation systems. Studies integrate clinical guidelines into recommendation algorithms underscore the value of combining general medical knowledge with specialized insights \cite{Zhang2022Knowledge-Enhanced,Gong2020SMR,Lu2019Incorporating}. Their work provides a foundation for HiRMed's dual-layer knowledge base architecture, which balances breadth and depth in medical reasoning. This dual-layer structure not only facilitates broad initial assessments but also enables precise, context-aware recommendations in specialized domains such as cardiology or endocrinology. By embedding both general and department-specific medical knowledge into a unified framework, such systems can dynamically adapt to the requirements of diverse clinical scenarios.

In summary, while existing systems have made strides in applying hierarchical models, RAG, and knowledge integration to healthcare, gaps remain in their ability to perform dynamic, context-aware medical reasoning. HiRMed addresses these gaps by synthesizing hierarchical tree-based structures, RAG-enhanced reasoning, and memory mechanisms, offering a novel approach to medical test recommendation.

\section{Methodology}

\subsection{Dataset and Knowledge Base}

Our system focuses on recommending medical tests in an outpatient setting, leveraging both structured patient data and comprehensive medical knowledge. To facilitate Retrieval-Augmented Generation (RAG), we compile an extensive dataset of outpatient visits and construct layered knowledge bases that integrate both general and department-specific medical information.

\subsubsection{Dataset}

We collect outpatient visit records from multiple hospital departments. Each record captures the patient's initial consultation, including physical parameters and presenting symptoms, as well as the diagnostic tests recommended by the physician. Over the course of treatment, additional diagnostic tests may also be performed, and these are recorded in our dataset alongside subsequent physician notes, final diagnoses, and treatment outcomes. By documenting both the initially recommended tests and any supplementary tests administered later, our dataset provides a comprehensive view of each patient's diagnostic and treatment pathway.

All outpatient records are preprocessed by removing personally identifiable information and standardizing key medical terms to ensure consistency across various hospital information systems. This results in a structured dataset containing fields such as patient demographics, reported symptoms, recommended diagnostic tests, follow-up tests, and corresponding clinical outcomes.

\subsubsection{Knowledge Base Construction}

We build a two-tiered knowledge base to enable hierarchical reasoning:
\begin{enumerate}
    \item \textit{Knowledge Base for Departments.} This repository covers broad, department-level medical knowledge (e.g., cardiology, endocrinology). Entries include standard disease--symptom correlations, commonly prescribed tests, and high-level clinical guidelines that apply to each specialty.
    \item \textit{Knowledge Base for Department Testing Items.} This repository contains more granular information regarding individual diagnostic tests (e.g., indications, interpretations, normal ranges). Each department has a specialized subset, allowing more precise recommendations for specific conditions.
\end{enumerate}

All textual knowledge in these repositories is embedded into a high-dimensional vector space using an OpenAI embedding model (e.g., \texttt{text-embedding-ada-002}). The embeddings are indexed in a FAISS-based vector database, which supports rapid similarity searches. This embedding process enables the system to dynamically retrieve relevant content based on a patient’s query, thus powering the retrieval-augmented generation steps at each stage of the recommendation pipeline.

\subsection{Model Selections}

Our hierarchical recommendation system incorporates three central models to handle vector representations, natural language reasoning, and final weighting of test suggestions. Each model is selected for its strengths in a particular aspect of the RAG workflow.

\subsubsection{Embedding Model}

We employ OpenAI’s Embedding API to convert both patient queries and knowledge-base text into numerical vectors. This transformation process captures nuanced semantic relationships, which is especially crucial for retrieving domain-relevant content in medical settings. The resulting embeddings are of manageable dimensionality, which allows for efficient similarity searches in the FAISS vector database. As a result, our system can swiftly locate high-value records or guidelines that match the patient’s symptoms and history.

\subsubsection{LLM API (GPT-O1)}

The GPT-O1 large language model serves as the core reasoning engine. It processes the text retrieved from the vector database, aligns it with the patient's context, and formulates initial or refined recommendations at each layer of the hierarchy. Through multi-turn interactions and dynamic prompts, GPT-O1 can generate intermediate diagnostic hypotheses and produce human-readable explanations that shed light on its reasoning. These outputs are instrumental in constructing a more interpretable and coherent test recommendation process.

\subsubsection{Weight Model (Fine-tuned LLaMA3.2-3B)}

We further enhance our recommendation pipeline with a fine-tuned version of LLaMA3.2-3B, which assigns weights or priorities to each recommended diagnostic test. The fine-tuning is conducted on historical outpatient data where each test has an associated physician-annotated relevance score. By learning from these annotations, the weight model can factor in patient demographics, symptom severity, and known comorbidities when ranking the final recommendations. This weighting mechanism is critical in distinguishing high-priority tests from those that may be secondary or less immediately necessary.

\subsection{Model Architecture}

\label{sec:model_architecture}

\begin{figure*}[!t]
    \centering
    \begin{subfigure}[b]{0.45\textwidth}
        \centering
        \includegraphics[width=\textwidth]{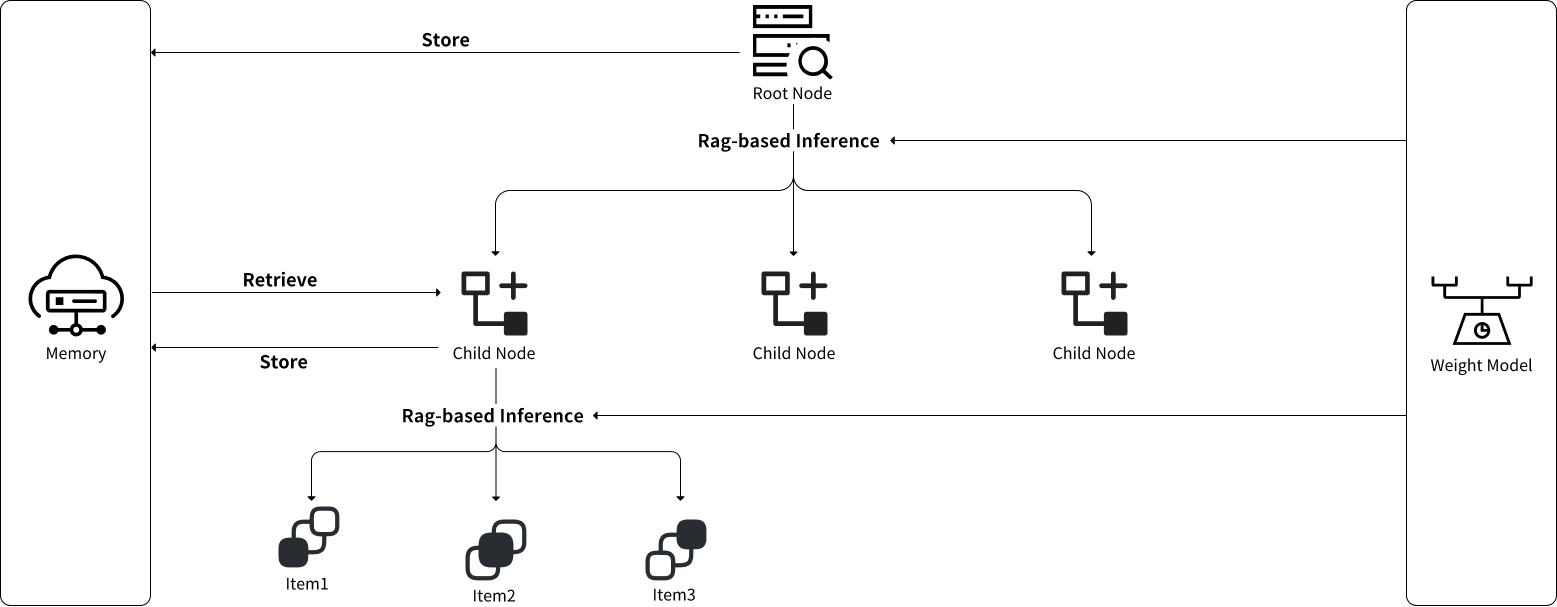}
        \caption{HiRMed System Architecture}
        \label{fig:pic1}
    \end{subfigure}
    \hfill
    \begin{subfigure}[b]{0.45\textwidth}
        \centering
        \includegraphics[width=\textwidth]{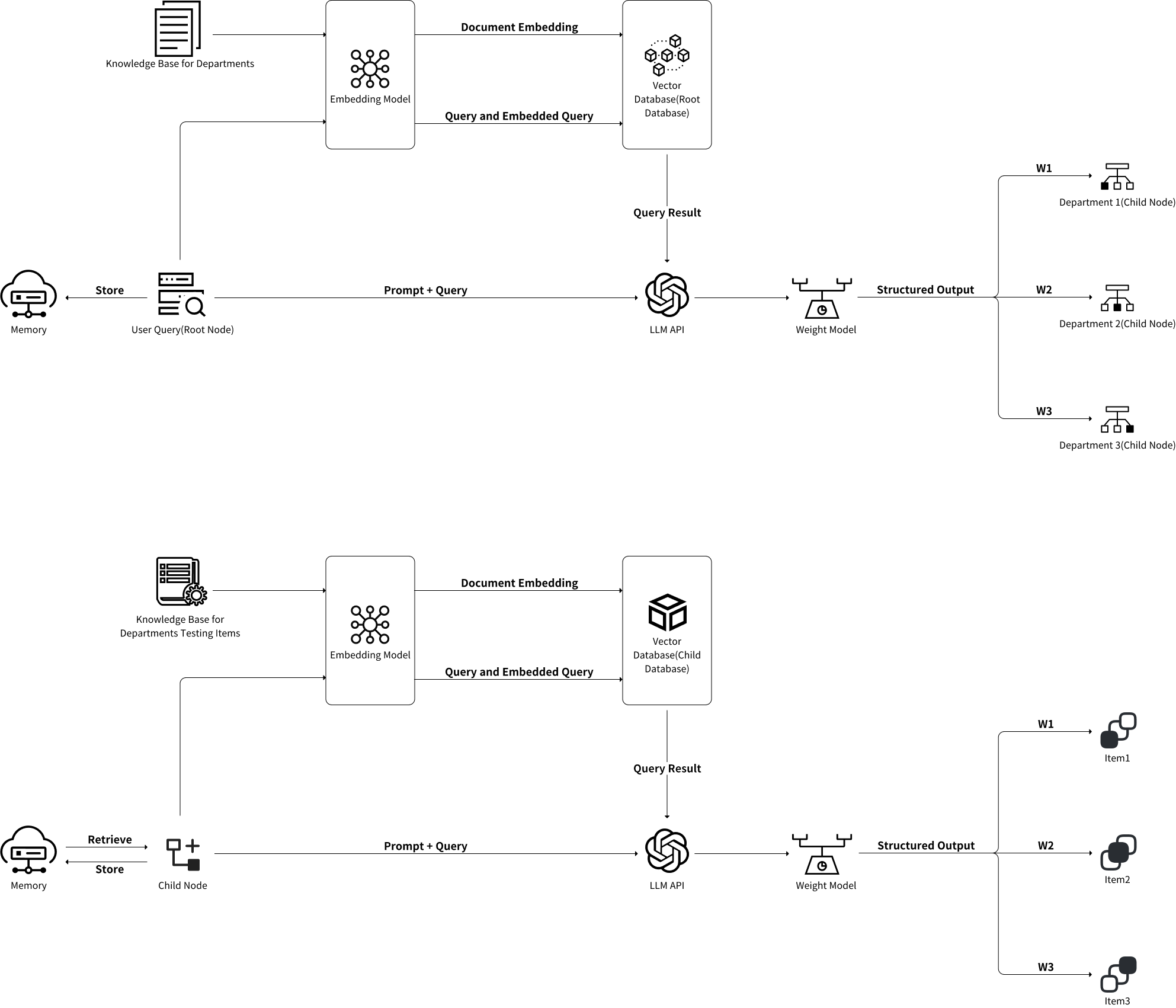}
        \caption{HiRMed Workflow}
        \label{fig:pic2}
    \end{subfigure}
    \caption{HiRMed System Overview}
    \label{fig:combined}
\end{figure*}

Figure~\ref{fig:pic1} illustrates the three-layer hierarchical architecture of our RAG-based medical test recommendation system. The design emulates real-world diagnostic logic by transitioning from broad, department-level considerations to specific test recommendations.

\begin{table}[!t]
\centering
\caption{Hierarchical Architecture Components and Their Functions}
\label{tab:hierarchy_functions}
\begin{tabular}{|p{0.15\linewidth}|p{0.35\linewidth}|p{0.35\linewidth}|}
\hline
\textbf{Layer} & \textbf{Key Components} & \textbf{Processing Functions} \\
\hline
Root & 
\begin{itemize}
\item Knowledge Base for Departments
\item Embedding Model
\item Vector Database
\item User Query Node
\end{itemize} & 
\begin{itemize}
\item Document Embedding
\item Query and Embedded Query Processing
\item Memory Storage and Retrieval
\item LLM-based Analysis
\end{itemize} \\
\hline
Department & 
\begin{itemize}
\item Department-specific Knowledge Base
\item Child Node Processing
\item Vector Database
\item Memory Management
\end{itemize} & 
\begin{itemize}
\item Specialized Knowledge Retrieval
\item Department-specific Query Processing
\item Weight Model Application
\item Structured Output Generation
\end{itemize} \\
\hline
Item & 
\begin{itemize}
\item Memory System
\item Root/Child Nodes
\item Process Units
\item Weight Model
\end{itemize} & 
\begin{itemize}
\item Flag-based Inference
\item Item-level Processing
\item Memory Storage/Retrieval
\item Final Result Generation
\end{itemize} \\
\hline
\end{tabular}
\end{table}

\subsubsection{Root Layer: Initial Analysis and Department Routing}

In the Root layer, the system takes the patient’s query (symptoms and other relevant data) and encodes it into a vector representation using the chosen embedding model. The vector database is then queried for semantically related information from the department-level knowledge base. GPT-O1 processes the retrieved documents alongside the patient context to identify potential diagnostic categories or specialties that should be explored. The fine-tuned LLaMA3.2-3B model subsequently assigns weights to each recommended specialty, allowing the system to rank and prioritize which departments may require more immediate attention.

\subsubsection{Department Layer: Specialty-Specific Reasoning}

Based on the Root layer’s output, the system transitions to the Department layer. Here, it retrieves domain-specific knowledge relevant to the selected specialty. GPT-O1 refines the diagnostic hypotheses by leveraging specialized guidelines, common presenting complaints, and department-focused protocols. This refined reasoning stage often involves narrowing down potential tests from a general list to those that align more closely with the patient’s clinical picture in the chosen specialty. LLaMA3.2-3B then updates the test prioritization, considering the context of that department’s typical patient risks and medical standards.

\subsubsection{Item Layer: Final Test Recommendations and Weighting}

In the final Item layer, a memory component consolidates the intermediate decisions and flagged symptoms to ensure continuity. GPT-O1 performs item-level reasoning, identifying the most appropriate diagnostic tests and resolving any inconsistencies or redundancies that might arise from previous layers. The fine-tuned LLaMA3.2-3B model then assigns definitive weights, reflecting the urgency or clinical utility of each test within the broader diagnostic strategy. The system ultimately produces a ranked list of recommended tests, accompanied by interpretive scores that highlight the underlying reasoning and urgency.

\subsubsection{Benefits of the Hierarchical Strategy}

Breaking down the recommendation process into these three layers enables more targeted retrieval, clearer interpretability, and better alignment with the medical decision-making process. The Root layer effectively broadens the diagnostic net, the Department layer introduces specialist insight, and the Item layer ensures that final test recommendations are both clinically focused and context-aware. The multi-layered approach, combined with a dedicated memory component and weighting mechanism, provides an adaptive framework that accommodates the complexity of real outpatient scenarios while maintaining transparent and consistent medical reasoning.

Table~\ref{tab:hierarchy_functions} presents a concise summary of the key activities and outputs at each layer of our hierarchical RAG-enhanced system.

\section{Experiments \& Results}

\subsection{Experimental Setup}

We evaluated HiRMed using a comprehensive clinical dataset comprising 125,000 outpatient visits across multiple hospital departments. The core component of our system - the preference model based on fine-tuned LLaMA3.2-3B - was trained to predict diagnostic test relevance scores by incorporating patient demographics, symptom severity, and medical history. The training process involved standardizing medical terminology, establishing symptom-test-outcome relationships, and preserving critical clinical decision features while removing personally identifiable information.
The system's knowledge base contains approximately 50,000 clinical guidelines, research papers, and standardized protocols, structured into two tiers: department-level knowledge (30\%) covering clinical guidelines and specialty-specific best practices, and test-specific knowledge (70\%) containing detailed testing parameters, indications, and interpretation guidelines. We focused our evaluation on three key departments - cardiology, endocrinology, and gastroenterology - as these specialties frequently encounter complex diagnostic scenarios requiring multiple tests and offer standardized clinical pathways for system validation.

\subsection{Baselines}

We compared HiRMed against several baseline approaches:

1) Traditional Vector Similarity (TVS): A standard retrieval-based system using cosine similarity between symptom embeddings and test descriptions.

2) Flat-RAG: A single-layer RAG system that directly maps symptoms to test recommendations without hierarchical reasoning.

\subsection{Evaluation Metrics}

We assessed system performance using the following metrics:

1) Coverage Rate (CR): The proportion of relevant diagnostic tests included in the recommendations.

2) Accuracy: The percentage of recommended tests that were deemed appropriate and necessary by reviewing physicians, measured by comparing system recommendations against expert-validated test orders in our ground truth dataset.

3) Miss Rate (MR): The proportion of critical tests (as determined by physician review) that were not recommended.

4) Clinical Relevance Score (CRS): Expert-assigned scores (1-5) evaluating the medical appropriateness of recommendations.

\subsection{Results}

\subsubsection{Overall Performance Analysis}

Table~\ref{tab:performance_comparison} demonstrates HiRMed's superior performance across all evaluation metrics compared to baseline approaches. The system achieved a coverage rate of 92.3\%, significantly outperforming both Flat-RAG (84.7\%) and TVS (72.8\%). In terms of accuracy, HiRMed reached 88.7\%, showing substantial improvement over Flat-RAG (82.4\%) and TVS (71.5\%). Most notably, HiRMed maintained a remarkably low miss rate of 2.1\% for critical tests, compared to 5.8\% for Flat-RAG and 10.6\% for TVS. The system's clinical relevance score of 4.3 out of 5 further validates its effectiveness, surpassing both Flat-RAG (3.7) and TVS (3.2).


\begin{table}[h]
\centering
\caption{Performance Comparison of Different Methods}
\label{tab:performance_comparison}
\begin{threeparttable}
\begin{tabular}{|l|c|c|c|c|}
\hline
\textbf{Method} & \textbf{Coverage Rate} & \textbf{Accuracy} & \textbf{Miss Rate} & \textbf{CRS} \\
\hline
HiRMed & \textbf{92.3} & \textbf{88.7} & \textbf{2.1} & \textbf{4.3} \\
Flat-RAG & 84.7 & 82.4 & 5.8 & 3.7 \\
TVS & 72.8 & 71.5 & 10.6 & 3.2 \\
\hline
\end{tabular}
\begin{tablenotes}
\item The first three columns (\textit{Coverage Rate}, \textit{Accuracy}, and \textit{Miss Rate}) are represented as percentages. The last column (\textit{CRS}) is the Clinical Relevance Score on a 1-5 scale.
\end{tablenotes}
\end{threeparttable}
\end{table}

\subsubsection{Department-Specific Analysis}

As shown in Table~\ref{tab:department_performance}, HiRMed maintained consistent performance across different medical departments while showing some variation in effectiveness. Cardiology demonstrated the strongest results with a 94.2\% coverage rate, 90.5\% accuracy, and a minimal miss rate of 1.8\%. Endocrinology and Gastroenterology showed similarly robust performance, with coverage rates above 90\% and accuracy rates exceeding 87\%. The consistent performance across departments underscores the system's adaptability to different medical specialties, with particularly strong results in cardiology due to the well-structured nature of cardiac diagnostic protocols.

\begin{table}[H]
\centering
\begin{threeparttable}
\caption{Department-Specific Performance Analysis}
\label{tab:department_performance}
\begin{tabular}{|l|c|c|c|}
\hline
\textbf{Department} & \textbf{Coverage Rate} & \textbf{Accuracy} & \textbf{Miss Rate} \\
\hline
Cardiology & \textbf{94.2} & \textbf{90.5} & \textbf{1.8} \\
Endocrinology & 91.7 & 88.3 & 2.2 \\
Gastroenterology & 90.8 & 87.4 & 2.4 \\
\hline
\end{tabular}
\begin{tablenotes}
\item All values in the columns are represented as percentages.
\end{tablenotes}
\end{threeparttable}
\end{table}

\subsubsection{Clinical Validation}

The system's effectiveness was further validated through rigorous clinical evaluation. A panel of 12 experienced clinicians reviewed 500 randomly selected cases from the test set, resulting in the high clinical relevance score of 4.3/5.0 shown in Table~\ref{tab:performance_comparison}. Clinicians particularly noted the system's ability to maintain coherent diagnostic pathways and effectively handle complex, multi-symptom cases.

\subsubsection{Component Analysis}

Table~\ref{tab:ablation_study} presents the results of our ablation studies, revealing the critical role of each system component. Removing the memory augmentation resulted in notable performance decreases across all metrics, with the coverage rate dropping by 8.2 percentage points and accuracy declining by 7.8 percentage points. The impact was even more pronounced when removing the department layer, causing an 11.6\% decrease in coverage rate and a 10.9\% decrease in accuracy. The most significant impact came from switching to a single knowledge base, resulting in a 13.7\% drop in coverage rate, a 12.8\% decrease in accuracy, and a 15.3\% increase in miss rate. These findings emphasize the importance of both the hierarchical structure and memory augmentation in maintaining the system's high performance levels.

\begin{table}[H]
\centering
\caption{Ablation Study Results}
\label{tab:ablation_study}
\begin{tabular}{|l|c|c|c|}
\hline
\textbf{System Variant} & \multicolumn{3}{c|}{\textbf{Performance Drop (\%)}} \\
\hline
& \textbf{Coverage Rate} & \textbf{Accuracy} & \textbf{Miss Rate} \\
\hline
Full HiRMed & 0.0 & 0.0 & 0.0 \\
w/o Memory & -8.2 & -7.8 & +9.4 \\
w/o Department Layer & -11.6 & -10.9 & +13.8 \\
Single Knowledge Base & -13.7 & -12.8 & +15.3 \\
\hline
\end{tabular}
\end{table}

\section{Conclusion}

This paper presented HiRMed, a novel hierarchical RAG-enhanced system for medical test recommendations that successfully addresses several key challenges in automated diagnostic support. Inspired by rigorous reasoning approaches like Chain-of-Thought and Monte Carlo Tree Search in game theory, we demonstrate that hierarchical structures can significantly enhance conventional reasoning tasks. By integrating a tree-structured architecture with RAG-based reasoning and memory augmentation, our system shows that decomposing complex decisions into hierarchical steps can achieve substantial improvements over traditional approaches across multiple performance metrics.

The experimental results validate our hierarchical approach, with HiRMed achieving a coverage rate of 92.3\% and an accuracy of 88.7\%, substantially outperforming baseline methods. Particularly noteworthy is the system's low miss rate of 2.1\% for critical tests, addressing a crucial concern in medical diagnosis. The consistent performance across different medical departments, with accuracy rates exceeding 87\% in all specialties, demonstrates that hierarchical reasoning can effectively handle diverse and complex scenarios.

Our ablation studies revealed the essential nature of each architectural component, particularly highlighting the importance of the dual-layer knowledge base and memory augmentation mechanism. The significant performance degradation observed when removing these components confirms that structured, hierarchical reasoning is crucial for maintaining high accuracy and comprehensive coverage in recommendation tasks. Furthermore, the positive clinical validation results, with a Clinical Relevance Score of 4.3/5, underscore that hierarchical decomposition of complex decisions can achieve practical utility in real-world settings.

Looking forward, this work opens several promising directions for future research in hierarchical reasoning systems. First, the architecture could be extended to incorporate real-time learning from physician feedback, allowing continuous refinement of the hierarchical decision process. Second, the system could be adapted to handle more complex scenarios involving multiple comorbidities and rare disease presentations. Finally, investigating the integration of temporal patient data could enhance the system's ability to perform multi-scale hierarchical reasoning across different time horizons.

HiRMed represents a significant step forward in both medical test recommendation systems and hierarchical reasoning approaches, demonstrating that combining tree-structured architectures with RAG-enhanced reasoning can effectively bridge the gap between general knowledge and specialized requirements. The system's success in maintaining both high accuracy and low miss rates, while providing interpretable recommendations, suggests that hierarchical reasoning could be a powerful paradigm for improving complex decision-making systems across various domains beyond healthcare.

\printbibliography

\end{document}